\documentclass[11pt]{article}
\usepackage{latexsym}
\newcommand{\beq}{\begin{equation}} 
\newcommand{\eeq}{\end{equation}}
\newcommand{\beqs}{\begin{eqnarray}} 
\newcommand{\eeqs}{\end{eqnarray}}
\newcommand{\tr}{\mathrm{tr}}
\newcommand{\WW}{{\cal W}}
\begin{document}
\begin{titlepage}
\vskip 2.5cm
\begin{center}
{\LARGE \bf Baryonic Corrections to Superpotentials}\\
\smallskip
{\LARGE \bf from Perturbation Theory}\\
\vspace{2.71cm}
{\Large
Riccardo Argurio,
Vanicson L. Campos, \\ 
\smallskip
Gabriele Ferretti and 
Rainer Heise}
\vskip 0.7cm
{\large \it Institute for Theoretical Physics - G\"oteborg University and \\
\smallskip
Chalmers University of Technology, 412 96 G\"oteborg, Sweden}
\vskip 0.3cm
\end{center}
\vspace{3.14cm}
\begin{abstract}
We study the corrections induced by a baryon vertex 
to the superpotential of SQCD with gauge group $SU(N)$ and $N$ quark flavors.
We first compute the corrections order by order using a standard field 
theory technique and derive the corresponding glueball superpotential
by ``integrating in'' the glueball field. The structure of the corrections
matches with the expectations from the recently introduced perturbative 
techniques. 
We then compute the first non-trivial contribution using this new technique
and find exact quantitative agreement. This involves cancellations between
diagrams that go beyond the planar approximation. 
\end{abstract}

\end{titlepage}

Important progress has recently been made in computing the glueball
superpotentials for a very large class of ${\cal N}=1$ gauge theories.
In~\cite{DV}, Dijkgraaf and Vafa have proposed a simple technique 
for computing the effective superpotential for the glueball
field $S = -{1\over 32 \pi^2}\mathrm{tr} \WW^\alpha \WW_\alpha$ by using a 
matrix model. 
This conjecture has more recently been proven in the 
work of~\cite{proof} where it was shown that the matrix model technique
gives the right result for a class of theories admitting a planar
expansion a la 't Hooft~\cite{thooft}. Namely, for theories with adjoint
matter, only planar (closed) diagrams contribute.

The conjecture can be extended to include matter in the fundamental
representation \cite{us}. As long as there are fields in the adjoint
the topological expansion is still possible, taking also
into account planar diagrams with one boundary.

However, the work of~\cite{proof} went
further than the matrix model analogy 
and gave a general expression for the glueball 
superpotential in terms of a field theory of chiral superfields even in the
case where the usual planar expansion is invalid.

It is the scope of this note to use and test this more general formula
in the physically interesting case of SQCD (i.e. without adjoint matter).
The first non-trivial interaction that can be added to the bare
superpotential is a baryon vertex
in a ${\cal N}=1$ gauge
theory with gauge group $SU(N_c)$ and $N_f=N_c\equiv N$ quark flavors.
Although the structure of the interaction does not allow for a planar
expansion we show, to first non-trivial order, that a cancellation similar
to the one taking place in~\cite{proof} also occurs here pointing to a 
larger range of applicability of the new techniques.

We consider a $SU(N)$ gauge theory with $N$ quark flavors and a tree level
superpotential
\beq
    W_{tree} = m \tr Q\tilde Q + b \det Q + \tilde b \det \tilde Q.
\eeq
The quark fields $Q$ and $\tilde Q$ 
are considered as $N\times N$ matrices
and given the same bare mass $m$. The coupling constants of the baryon vertices
are denoted by $b$ and $\tilde b$. 
The meson and baryon operators are thus
\beq
     M^j_i= Q^a_i\tilde Q^j_a, \quad 
     B = \det Q, \quad \tilde B = \det \tilde Q.
\eeq
Before adding $W_{tree}$,
the theory has $N$ massless quarks and holomorphic 
scale $\hat \Lambda$. At the quantum level, there is no effective
superpotential but the moduli space receives corrections~\cite{moduli}.
After we add $W_{tree}$ and integrate out the massive fields, 
the theory at low energies 
is pure Super Yang-Mills with holomorphic scale $\Lambda$
given by $\Lambda^3 = m \hat \Lambda^2$. 

It is easy to obtain the corrections to the superpotential of the low energy
SYM theory by integrating out the composite fields.
First we impose the condition on the quantum moduli space~\cite{moduli}
by writing the superpotential of the high energy theory with the help of a 
Lagrange multiplier $\xi$:
\beq
    W_{high} = \xi ( \hat\Lambda^{2N}-\det M + B \tilde B).
\eeq
Then we integrate out $M$, $B$ and $\tilde B$ (and $\xi$) from:
\beq
    W_{tot} = W_{high} + W_{tree}.
\eeq
to find that $\xi$ obeys the constraint:
\beq
    \hat\Lambda^{2N} -\left(\frac{m}{\xi}\right)^{N \over N-1} + 
\frac{b \tilde b}{\xi^2}     =0. \label{constr}
\eeq
The effective superpotential for the low energy theory is now:
\beq
    W_{low} = N \xi \hat\Lambda^{2N} +(N-2) 
              \frac{b \tilde b}{\xi}. \label{low}
\eeq

The constraint (\ref{constr}) can be solved algebraically for small 
$N$s,\footnote{
For instance, for $SU(2)$. Note that this case is trivial
since the baryonic interaction is just an off-diagonal mass term, the
${\bf 2}$ and ${\bf \bar 2}$ representations being equal. 
Nevertheless if we insist in expanding in $b \tilde b$ all
the following formulas hold for this case, and match the 
perturbative derivation, even if all the terms are proportional to $S$.}
but in general we must content ourselves with a series expansion in
powers of $b\tilde b$: 
\beq
    W_{low} = N \Lambda^3 \left(1 - \frac{1}{N}t -
       \frac{N-1}{2 N^2}t^2 - \frac{(N-1)(4N-5)}{6N^3}t^3 +\dots \right)
    \label{lown}
\eeq
where we have used the dimensionless and chargeless variable
\beq
     t = \frac{\hat\Lambda^{2N-4}b\tilde b}{m^2}={\Lambda^{3(N-2)}b\tilde b
\over m^N}.
\eeq

To compare with the results of~\cite{DV} and~\cite{proof} we need to 
integrate in~\cite{integratein} the glueball field to obtain,
for $\beta = \frac{b \tilde b}{m^N}$:
\beqs
    W_{glue} &=& N S (-\log(S/\Lambda^3) + 1) \label{glue} \\ & & \nonumber
 - \beta S^{N-1} -
        \frac{N-1}{2} \beta^2 S^{2N-3} - 
        \frac{(N-1)(3N-4)}{6}\beta^3 S^{3N-5} + \dots 
\eeqs

One can already notice that the functional dependence of $W_{glue}$ on all
the variables is in agreement with the counting of~\cite{proof}, that is the 
contribution from diagrams with $2k$ baryon vertices will carry $Nk$ 
propagators and require $Nk - 2k + 1$ factors of $S$ to soak up the
fermionic zero modes. The generic form of the numerical coefficients is also 
worth commenting on -- note that there appear no powers of $N$ in the 
denominator (contrary to (\ref{lown})) pointing possibly to a  
simple counting argument even in the absence of a planar expansion.

We will now reproduce the first coefficient with the technique of~\cite{proof}
showing that we indeed find an amplitude which is topological in nature.
However this comes as a result of a subtle cancellation between  
diagrams which would naively be thought as contributing to different 
orders in the large $N$ expansion. We stress that our computation
is a purely field theoretic one.

We thus compute the effective action of a theory
with matter fields in a background gaugino condensate $S=-{1\over 32\pi^2}
\tr \WW^\alpha \WW_\alpha$, following
\cite{proof} (and extending to complex representations of the gauge group), as:
\beq
e^{-\int d^4x\, d^2\theta\, W_{eff}(S)}=\int {\cal D}Q\, {\cal D}\tilde Q\, 
e^{-\int d^4x\, d^2\theta\, [-Q(\Box -i{\cal W}^\alpha D_\alpha -m)\tilde Q
+W_{int}(Q,\tilde Q)]} , \label{path}
\eeq
where $W_{int}(Q,\tilde Q)$ contains the interaction part of the tree level
superpotential. It is worth noting that as far as the effective superpotential
is concerned, the full path integral becomes holomorphic. This is proven 
in \cite{proof}, and we refer to that paper for more details.
From the free part of the action in (\ref{path}) we can read off the
propagator, which in momentum space is:
\beq
\Delta(p,\pi)^{ai}_{bj}=
\int_0^\infty ds 
\left[e^{-s(p^2+m+{\cal W}^\alpha \pi_\alpha)}\right]^a_b \delta^i_j.
\label{prop}
\eeq
Note that we have taken for simplicity
a mass matrix proportional to the identity. We can always restore generality
at the end of the computation using symmetry arguments.

If now we take $W_{int}=b\det Q +\tilde b \det \tilde Q$, we have:
\beq
e^{-\int d^4x\, d^2\theta\, W_{eff}(S)}= \langle e^{-\int d^4x\, d^2\theta\, 
(b\det Q +\tilde b \det \tilde Q)}\rangle 
\eeq
so that:
\beqs W_{eff}(S) 
&=& -\sum_{n=1}^\infty {(b\tilde b)^n \over (n!)^2 }\int
d^4x_1\, d^2\theta_1\, \dots d^4x_{2n-1}\, d^2\theta_{2n-1}\, \times 
\\ & & \times
 \langle \det Q(z_1) \dots \det Q(z_n) \nonumber
\det \tilde Q(z_{n+1})\dots \det \tilde Q(0) \rangle_c
\eeqs
where only connected amplitudes are computed, and $z$ represents a
(chiral) superspace coordinate.

The first term in the expansion, which we are going to compute, 
is given by $(-b\tilde b)$ times the $(N-1)$-loop amplitude:
\beqs
{\cal A}&=&\int d^4x\, d^2\theta\, \langle \det Q(z)
\det \tilde Q(0)\rangle \label{ampli}\\
&=& \int d^4x\, d^2\theta\, \epsilon_{a_1\dots a_N}\epsilon^{b_1\dots b_N}
\langle Q^{a_1}_1 \dots Q^{a_N}_N(z)\tilde Q^1_{b_1}\dots
\tilde Q^N_{b_N}(0) \rangle \nonumber
\eeqs
We can now insert the propagators (\ref{prop}). Note that they connect
only the same flavors, 
so that the symmetry factor of the diagram is just unity.
The integrals over $x$ and $\theta$ are going to enforce the conservation
of momentum $\sum p_i=0$ and likewise $\sum \pi_i=0$.

The integral on the remaining bosonic momenta then gives:
\beq
\int{d^4p_1 \over (2\pi)^4}\dots {d^4p_{N-1} \over (2\pi)^4}
e^{-s_1 p_1^2-\dots -s_{N-1}p_{N-1}^2 -s_N(p_1+\dots+p_{N-1})^2} =
{(4\pi)^{-2(N-1)}\over (\det M(s))^2}, 
\eeq
where:
\beq
\det M(s)=s_1\dots s_{N-1}+s_1\dots s_{N-2}s_N +\dots + s_2\dots s_N.
\label{deter}
\eeq
We are also left with $N-1$ integrals on the grassmannian momenta $\pi_i$.
The simple treatment for matter in the adjoint given in \cite{proof}
is not applicable to our problem because of its intrinsic non-planarity.
We have thus to perform the integrals explicitly:
\beq
\int d^2\pi_1 \dots d^2\pi_{N-1}\epsilon_{a_1\dots a_N}\epsilon^{b_1\dots b_N}
\left[e^{-s_1\WW^\alpha {\pi_1}_\alpha}\right]^{a_1}_{b_1} \dots
\left[e^{s_N \WW^\alpha (\pi_1+\dots +\pi_{N-1})_\alpha}\right]^{a_N}_{b_N}.
\label{afermi}
\eeq
We start by expanding the grassmannian piece of the propagator (\ref{prop}):
\beq
\left[e^{-s\WW^\alpha \pi_\alpha}\right]^a_b=\delta^a_b-s(\WW^\alpha)^a_b
\pi_\alpha-s^2(\WW^2)^a_b \pi^2.
\eeq
We are integrating over $2N-2$ grassmannian variables a polynomial 
which is of degree $2N$. 

Now, if we write:
\beq
\epsilon_{a_1\dots a_N}\epsilon^{b_1\dots b_N}=
\delta^{b_1}_{a_1}\dots \delta^{b_N}_{a_N}\pm (\mbox{permutations of}\ b_i),
\eeq
we can easily see that the number of traces over the color indices
in a particular term
will be given by $N$ minus
the number of transpositions of the related permutation.

Since we take the gauge group to be $SU(N)$, we have trivially
that $\tr \WW^\alpha=0$. Furthermore, as far as the first correction of
the effective superpotential is concerned, we can 
assume that only $\tr \WW^2 
\propto S$ is non-zero, while all higher traces are taken to vanish,
as in \cite{proof}.
Note that this will not be true for higher order terms in $b \tilde b$:
for instance in $SU(3)$ the term $\tr \WW^2\WW^2\WW^2$ cannot be
set to zero independently from $\tr \WW^2$ and this will affect
the calculation of the second order correction. 

We thus see that the only terms contributing to the amplitude are the
following: the trivial permutation, which
leads to $N$ traces and thus is proportional to $N(\tr\WW^2)^{N-1}$,
and the single transpositions, which have $N-1$ traces and are thus
proportional to $(\tr\WW^2)^{N-1}$.
It is worth noting that these two contributions would arise at different
orders in the large $N$ expansion, if we were to count only the factors
of $N$ arising from the traces over the colors and not the symmetry
factors due to the vertex which is also of order $N$.

After performing the integrals over the grassmannian momenta, 
the term proportional to the trivial permutation gives:
\beq
N(-\tr\WW^2)^{N-1}(s_1^2\dots s_{N-1}^2+\dots +s_2^2\dots s_N^2).
\eeq
The terms with one transposition, after integration, contribute like:
\beq
-(-\tr\WW^2)^{N-1}\left[(s_1-s_2)^2s_3^2\dots s_N^2+\dots +
s_1^2\dots s_{N-2}^2(s_{N-1}-s_N)^2\right],
\eeq
that is, there are ${1\over 2}N(N-1)$ terms of the above form 
corresponding to every couple
in the $N$ Schwinger parameters.

Remarkably, summing these two different contributions gives 
the square of the determinant (\ref{deter}), which cancels exactly
the denominator obtained from integrating over the bosonic momenta.
Recalling that $S=-{1\over 16 \pi^2}\tr \WW^2$, we can write the
amplitude (\ref{ampli}) as:
\beq
{\cal A}=S^{N-1}\int ds_1 \dots ds_N\, e^{-m\sum s_i}={1\over m^N}
S^{N-1}.
\label{first}
\eeq
Restoring the notation of Eq.~(\ref{glue}), we have that:
\beq
W_{eff}(S)= N S (-\log(S/\Lambda^3) + 1)-\beta S^{N-1} + O(\beta^2),
\eeq
thus obtaining the first correction
to the pure SYM Veneziano-Yankielowicz glueball superpotential \cite{VY}.
 
The simplicity of the result (\ref{first}) should be confronted with
its non-triviality. The cancellation comes about between diagrams of
two different orders, the leading one giving a potential contribution
of order $NS^{N-1}$. 
However the fact that the final integral in (\ref{first}) reduces
to a topological one as in the theories exemplified in \cite{proof}
could indicate that some geometric picture is possible also in this
case. It would be interesting to push further and see if similar
cancellations arise also in higher order terms. This is currently
under investigation.

\section*{Acknowledgments}
We would like to thank G.~Bonelli, M.~Cederwall and P.~Salomonson for 
discussions. 
This work is partly supported by EU contract HPRN-CT-2000-00122.


\begin{thebibliography}{99}

\bibitem{DV}
R.~Dijkgraaf and C.~Vafa, 
``A Perturbative Window into Non-Perturbative Physics,''
arXiv:hep-th/0208048.




\bibitem{proof}
R.~Dijkgraaf, M.~T.~Grisaru, C.~S.~Lam, C.~Vafa and D.~Zanon,
``Perturbative computation of glueball superpotentials,''
arXiv:hep-th/0211017.

\bibitem{thooft}
G.~'t Hooft,
``A Planar Diagram Theory For Strong Interactions,''
Nucl.\ Phys.\ B {\bf 72} (1974) 461.

\bibitem{us}
R.~Argurio, V.~L.~Campos, G.~Ferretti and R.~Heise,
``Exact superpotentials for theories with flavors via a matrix integral,''
arXiv:hep-th/0210291;

J.~McGreevy,
``Adding flavor to Dijkgraaf-Vafa,''
arXiv:hep-th/0211009;

H.~Suzuki,
``Perturbative derivation of exact superpotential for meson fields from  
matrix theories with one flavour,''
arXiv:hep-th/0211052;

I.~Bena and R.~Roiban,
``Exact superpotentials in N = 1 theories with flavor and their matrix  
model formulation,''
arXiv:hep-th/0211075;

Y.~Demasure and R.~A.~Janik,
``Effective matter superpotentials from Wishart random matrices,''
arXiv:hep-th/0211082;

Y.~Tachikawa,
``Derivation of the Konishi anomaly relation from Dijkgraaf-Vafa with  
(bi-)fundamental matters,''
arXiv:hep-th/0211189;

B.~Feng,
``Seiberg Duality in Matrix Model,''
arXiv:hep-th/0211202;

B.~Feng and Y.~H.~He, 
``Seiberg Duality in Matrix Models II,''
arXiv:hep-th/0211234.

\bibitem{moduli}
N.~Seiberg,
``Exact results on the space of vacua of four-dimensional SUSY 
gauge theories,''
Phys.\ Rev.\ D {\bf 49} (1994) 6857
[arXiv:hep-th/9402044].

\bibitem{integratein}
K.~A.~Intriligator, R.~G.~Leigh and N.~Seiberg,
``Exact superpotentials in four-dimensions,''
Phys.\ Rev.\ D {\bf 50} (1994) 1092
[arXiv:hep-th/9403198].

K.~A.~Intriligator,
``'Integrating in' and exact superpotentials in 4-d,''
Phys.\ Lett.\ B {\bf 336} (1994) 409
[arXiv:hep-th/9407106].



\bibitem{VY}
G.~Veneziano and S.~Yankielowicz,
``An Effective Lagrangian For The Pure N=1 Supersymmetric Yang-Mills Theory,''
Phys.\ Lett.\ B {\bf 113} (1982) 231.


\end{thebibliography}
\end{document}